\documentclass[conference]{IEEEtran}
\usepackage{graphicx}
\usepackage{amsmath, amssymb}
\usepackage{algorithm}
\usepackage{algorithmic}
\usepackage{hyperref}
\usepackage{listings}
\usepackage{xcolor}
\usepackage{listings}
\usepackage{amsmath}
\usepackage{amsfonts}
\usepackage{placeins}
\usepackage{tabularx}
\usepackage{ltxtable}
\usepackage{multirow}
\usepackage{float}
\usepackage{tcolorbox}

\lstdefinelanguage{EBNF}{
    keywords={},
    sensitive=false,
    comment=[l]{//},
    morestring=[b]",
}
\lstdefinelanguage{PDDL}{
    keywords={define, domain, types, predicates, action, parameters, precondition, effect, and, or, not},
    sensitive=false,
    comment=[l]{;},
    morestring=[b]",
}
\begin{document}

\title{Bridging Language Models and Formal Methods for Intent-Driven Optical Network Design}

\author{
    \IEEEauthorblockN{Anis Bekri\IEEEauthorrefmark{1}\IEEEauthorrefmark{2}, 
                      Amar Abane\IEEEauthorrefmark{2}, 
                      Abdella Battou\IEEEauthorrefmark{2},
                      Saddek Bensalem\IEEEauthorrefmark{1}}              
    \IEEEauthorblockA{\IEEEauthorrefmark{1}Université Grenoble Alpes, Grenoble, France}
    \IEEEauthorblockA{\IEEEauthorrefmark{2}National Institute of Standards and Technology, Gaithersburg, USA}
    \IEEEauthorblockA{Email: anis.bekri@nist.gov, amar.abane@nist.gov, abdella.battou@nist.gov, saddek.bensalem@univ-grenoble-alpes.fr}
}

\maketitle
\begin{abstract}
Intent-Based Networking (IBN) aims to simplify network management by enabling users to specify high-level goals that drive automated network design and configuration. However, translating informal natural-language intents into formally correct optical network topologies remains challenging due to inherent ambiguity and lack of rigor in Large Language Models (LLMs). To address this, we propose a novel hybrid pipeline that integrates LLM-based intent parsing, formal methods, and Optical Retrieval-Augmented Generation (RAG). By enriching design decisions with domain-specific optical standards and systematically incorporating symbolic reasoning and verification techniques, our pipeline generates explainable, verifiable, and trustworthy optical network designs. This approach significantly advances IBN by ensuring reliability and correctness, essential for mission-critical networking tasks.
\end{abstract}

\section{Introduction}

Intent-Based Networking (IBN) simplifies network management by allowing users to express high-level objectives—such as connectivity, performance, or security—without specifying implementation details~\cite{clemm2022intent, leivadeas2022survey}. Standardization bodies like TM Forum and the Internet Engineering Task Force define intent as a declarative statement of desired outcomes, delegating the detailed configuration and implementation tasks to automated systems. By abstracting away low-level complexities, IBN significantly reduces operational overhead, human error, and management complexity~\cite{leivadeas2022survey}.

Existing research predominantly explores intent translation into configurations or incremental topology adjustments~\cite{ifland2024genet, wang2023making}, but largely overlooks the initial phase of comprehensive network design, particularly for optical networks. Effective optical network design demands rigorous consideration of physical-layer constraints, including wavelength continuity, latency budgets, redundancy schemes, and equipment placement. Poor initial design decisions can lead to significant performance degradation or expensive reconfigurations throughout the operational lifecycle~\cite{velasco2022intent, wang2024large}. Incorporating automated design and planning into IBN frameworks is essential to realizing full lifecycle management in autonomous networks—from intent expression and design generation to deployment, monitoring, and optimization.

Recently, LLMs have demonstrated remarkable capabilities in interpreting complex natural-language requirements, opening new avenues for intent-based network automation in telecommunications~\cite{wu2024netllm, wang2024netconfeval, hang2024large}. However, LLMs exhibit several limitations that hinder their effectiveness for broader telecom applications. These limitations include difficulty capturing causal and mathematical relationships inherent in telecom systems, leading to plausible yet incorrect outputs—often termed hallucinations; a lack of rigorous mathematical foundations necessary for verifying and manipulating network parameters; reliance on static training data that fails to address dynamic and evolving telecom environments; and the absence of built-in regulatory guardrails, risking non-compliance with essential industry standards~\cite{shahid2025large}.

These inherent shortcomings are especially problematic in optical network design, which fundamentally constitutes a constraint satisfaction and planning problem involving rigorous compliance with network constraints such as wavelength continuity, latency bounds, redundancy feasibility, and budget requirements. Recent studies illustrate the limitations clearly; Kambhampati et al.~\cite{kambhampati2024position} report that only around 12\% of GPT-4-generated network plans are executable without formal constraints, significantly eroding trust in automation driven purely by LLMs.

To address these limitations, researchers have explored several complementary techniques to enhance the reliability of LLM-based approaches in telecom scenarios. Retrieval-Augmented Generation (RAG), which supplements LLM outputs with authoritative, domain-specific knowledge—such as telecom standards or vendor specifications—has shown promise in reducing inaccuracies and infeasibilities~\cite{bornea2024telcoragnavigatingchallengesretrievalaugmented}. Symbolic verification and constraint-solving approaches have also been employed to rigorously ensure correctness and enforce technical constraints within generated outputs~\cite{li2024formal}. Additionally, neuro-symbolic methods, which integrate the flexibility of LLMs with formal symbolic reasoning, have emerged as particularly promising for handling complex intent-driven network planning problems~\cite{li2024formal}.

In this paper, we present an \textit{intent-to-design copilot for optical networks}, aimed at accelerating and improving the design and planning stages of optical networks from high-level natural language intents. This copilot addresses the challenge of translating vague or underspecified user requirements into technically accurate, constraint-compliant, and deployment-ready network topologies.

Our solution is designed as a hybrid pipeline that combines the language flexibility and intent understanding capabilities of LLMs with the precision and correctness guarantees of symbolic verification methods. The pipeline also incorporates Optical RAG, a domain-specific retrieval system we designed to inject guidance from optical networks standards and vendor documentation. 
The pipeline operates through three integrated stages:
\begin{enumerate}
    \item Intent parsing and validation using LLMs with Context-Free Grammar (CFG) structural verification.
    \item Knowledge enrichment through Optical RAG integration with authoritative networking standards.
    \item Deployment planning via Planning Domain Definition Language (PDDL) for formal feasibility verification.
\end{enumerate}

To our knowledge, this is the first proposal that systematically integrates all three paradigms—LLM parsing, domain-specific RAG, and symbolic reasoning—to produce formally verified optical network designs from free-form natural language intents. This advances the automation of network architecture generation, reducing reliance on manual design cycles and increasing the trustworthiness of AI-driven network planning.

The rest of the paper is structured as follows. Section \ref{sec:related_work} reviews related work and background on IBN, LLM limitations, and hybrid methods. Section \ref{sec:architecture} describes the three-stage architecture of the proposed hybrid pipeline. Section \ref{sec:eval} presents implementation and case studies evaluation of our prototype and Section \ref{sec:concl} concludes with future directions.
\section{Background and Related Work}
\label{sec:related_work}

\subsection{Intent-Based Networking}

IBN has become a central paradigm in simplifying network management by enabling operators to specify high-level objectives, leaving the system to determine and enforce the necessary configurations. Traditional IBN systems relied on structured representations like JSON, YAML, or domain-specific languages to capture intent, often requiring users to understand the underlying data models. However, the emergence of LLMs has shifted the focus toward more natural interfaces that allow operators to express intents in plain language.

Recent studies have introduced various LLM-centric IBN frameworks aimed at automating intent lifecycle processes, such as decomposition, translation, activation, and assurance. Mekrache et al.~\cite{mekrache2024intent} propose a complete LLM-based architecture to manage the end-to-end intent lifecycle of 5G networks. Their framework demonstrates how LLMs can decompose high-level intents and activate corresponding configurations, reducing the barrier for non-expert users.

Similarly, Orlandi et al.~\cite{orlandi2024intent} present a modular architecture for managing 5G/6G services using NLP-powered interfaces, showing how user-friendly engagement channels (like chatbots) can drive configuration automation in network slicing contexts. Their architecture integrates standardized models to align user expectations with underlying orchestration mechanisms.

Ifland et al.~\cite{ifland2024genet} go a step further with GeNet, a multimodal LLM-powered co-pilot capable of understanding both textual and visual input to update network configurations and topologies. While GeNet shows strong potential in translating intent into meaningful topology edits and configuration synthesis, it operates within an assumed topology.

Other frameworks, such as NETBUDDY~\cite{wang2023making}, also explore using LLMs for generating and adapting network configurations from high-level policies. NETBUDDY decomposes the configuration process into structured stages and incorporates verification mechanisms to ensure correctness. 

While these efforts advance LLM-driven IBN, most address intent-to-configuration or modification tasks and presume that network design (topology, equipment, links) is already in place. This represents a fundamental limitation as configuration automation cannot address the crucial decisions of network architecture, equipment selection, and physical topology design, where early decisions (e.g., fiber type, redundancy, device placement) have long-term performance implications.
Our work fills this gap by proposing a pipeline that performs intent-to-design from scratch, or at the planning stage of an optical network.

\subsection{IBN for Optical Networks}
Applying IBN to optical networks introduces unique challenges due to strict physical-layer constraints such as wavelength continuity, signal regeneration, latency bounds, and protection requirements. Traditional IBN approaches, largely developed for packet-based systems, are often insufficient to meet the demands of optical infrastructure, which requires highly accurate modeling and control across the physical and logical layers.

Velasco et al.~\cite{velasco2022intent} provide a comprehensive treatment of this problem space. Their tutorial introduces an IBN framework adapted to optical networking, emphasizing the use of machine learning (ML) pipelines and data analytics for tasks such as Quality of Transmission (QoT) estimation, proactive self-configuration, and dynamic capacity adjustment. They propose an architecture that tightly couples intent interpretation with ML-based assurance loops and orchestration systems, and show how ML Function Orchestrators (MLFOs) can be used to deploy and reconfigure intent-driven analytics chains associated with optical network entities (e.g., lightpaths).

Building on the integration of AI into optical systems, Wang et al.~\cite{wang2024large} propose a conceptual framework for LLM-driven automation in optical networks. They introduce an AI agent embedded within the control plane that can interpret user intents, retrieve domain-specific knowledge through prompt engineering and RAG, and generate control instructions for tasks such as alarm analysis, lightpath optimization, and device configuration. Their framework demonstrates the viability of LLMs for physical layer control and task decomposition via prompting strategies like Chain-of-Thought (CoT) and Tree-of-Thought (ToT).

Despite these significant contributions, existing work remains primarily focused on operational intelligence, configuration, or automation within pre-defined infrastructures—assuming a fixed network topology. This leaves a critical gap in automating the early phases of the network lifecycle, such as equipment selection, topology generation, and constraint-aware layout planning, from high-level user intents.

\subsection{Challenges of LLM in Networking}

A fundamental limitation of LLMs is their susceptibility to hallucination—producing plausible but incorrect outputs. Xu et al.~\cite{xu2024hallucination} formally prove that hallucination is not just a practical issue but an unavoidable theoretical consequence of the computational limitations of LLMs. Their work demonstrates that for any computable LLM, there exist computable functions (e.g., real-world knowledge or rules) that the model cannot perfectly learn or reproduce. Therefore, inconsistencies between model outputs and the ground truth are not anomalies—they are inevitable, particularly in high-stakes domains like networking and telecom where factual correctness is critical.

In the networking context, these limitations are compounded by domain-specific challenges. The authors in~\cite{shahid2025large} emphasize that directly applying generic LLMs to telecom systems often fails due to their lack of mathematical grounding, poor handling of causal relationships, static knowledge base, and absence of domain-specific regulatory constraints. For example, standard LLMs may not understand how increasing transmission frequency affects propagation loss or how power and distance interact in wireless signal equations. These models often generate technically plausible answers without respecting the physical laws or telecom standards (e.g., 3GPP). This makes them inadequate for tasks like network planning, configuration validation, or KPI prediction, where mathematical reasoning, compliance, and dynamic adaptability are crucial.

To address these issues, the authors advocate for the development of Large Telecom Models (LTMs) that integrate causal reasoning, neuro-symbolic AI, and RAG techniques. Our work aligns with this vision by combining these complementary approaches specifically for optical network design.

\subsection{Mitigating LLM Limitations with Formal Methods and Hybrid Architectures}

Despite the growing utility of LLMs in networking and automation, their use in safety-critical or constraint-heavy domains like network design remains hindered by fundamental limitations. Chief among these are hallucinations, lack of planning capabilities, and the inability to guarantee formal correctness. Kambhampati et al.~\cite{kambhampati2024position} argue that LLMs cannot reliably perform planning or self-verification tasks, showing that fewer than 12\% of plans generated by GPT-4 are executable without further constraints. Instead, they propose LLM-Modulo Frameworks, where LLMs are paired with external verifiers that provide soundness checks and model-based reasoning capabilities.

A related direction leverages formal methods as external correctness oracles for LLM outputs. Jha et al.\cite{jha2023dehallucinating} present an architecture where formal logic specifications guide iterative prompting to correct hallucinated outputs from LLMs. Rather than trusting the initial LLM generation, their system interacts in a self-correcting loop until formal constraints are satisfied. Similarly, the Formal-LLM framework~\cite{li2024formal} constrains LLM-based planning using a pushdown automaton derived from a CFG, enforcing syntactic correctness in the generated plans. Recent work by Smirnov et al.~\cite{smirnov2024generating} explores generating consistent PDDL domains directly with LLMs, demonstrating the feasibility of LLM-assisted formal planning domain creation. These hybrid designs show that combining LLM flexibility with formal symbolic layers increases both accuracy and controllability.

Zhang et al.~\cite{zhang2024fusion} propose a comprehensive roadmap to merge LLMs with formal methods through a bi-directional integration: FMs-for-LLMs, where symbolic verifiers enhance trust and correctness in generation, and LLMs-for-FMs, where language models improve the usability and adaptability of formal verification tools. This mutual enhancement fosters trustworthy AI systems, particularly relevant for applications like network design where both user-facing natural language interfaces and rigorous backend validation are necessary.

Another powerful strategy to mitigate hallucinations is RAG. Rather than relying purely on internal model parameters, RAG supplements LLMs with dynamic access to domain-specific knowledge sources. In highly technical fields like telecom and networking, RAG systems have been shown to drastically reduce hallucinations and improve factual grounding. Bornea et al.~\cite{bornea2024telcoragnavigatingchallengesretrievalaugmented} introduce Telco-RAG, a domain-adapted RAG system optimized for telecom standard documents (3GPP). Their pipeline includes query refinement, vocabulary-based glossary enhancement, and a neural router to select only the most relevant document subsets, thus reducing memory usage and improving answer accuracy. Telco-RAG achieves substantial accuracy gains over baseline RAG systems and provides an open-source foundation for technical domain RAGs. The Optical RAG component in our proposed pipeline is adapted from Telco-RAG, tailored specifically to optical networking standards and equipment documentation.

\subsection{Summary}

While significant progress has been made in LLM-based intent interpretation and formal verification methods independently, the systematic integration of these approaches for comprehensive optical network design remains unexplored. Our work addresses this gap by proposing a hybrid framework that integrates structured intent parsing using CFG, domain-specific knowledge injection via RAG, and formal deployment verification through PDDL to generate complete, verified optical network designs from natural language specifications.
\section{Architecture Overview}
\label{sec:architecture}

Our approach views network intent-to-design as a structured planning problem. Therefore, we propose a hybrid architecture that unites the flexibility of natural language processing with the rigor of formal verification.
By modeling optical network design as a formal planning task, we enable the use of LLMs for natural language interaction while relying on symbolic reasoning for correctness and constraint satisfaction. This hybrid approach combines user accessibility with formal guarantees—capabilities that purely LLM-based systems cannot offer.

\begin{figure*}
    \centering
    \includegraphics[width=0.95\textwidth]{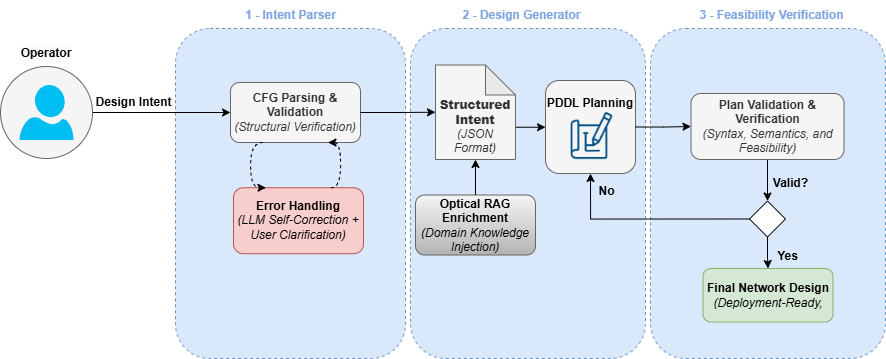}
    \caption{Intent-to-design pipeline showing three-stage architecture with formal validation at each level}
    \label{fig:pipeline-overview}
\end{figure*}

The architecture bridges the gap between ambiguous user input and deployable configurations, ensuring both structural and semantic validity.
Figure~\ref{fig:pipeline-overview} presents the three-stage pipeline, with each stage addressing a distinct layer of validation.

\subsection{Design Rationale for Component Selection}

\subsubsection{Context-Free Grammar (CFG) for Structural Validation}

We employ CFG parsing as the first line of defense against intent ambiguity and inconsistent formatting. While LLMs are effective at understanding semantics, they often produce inconsistent output structures for similar inputs. The CFG addresses this by enforcing structural disambiguation, standardizing the output format, and applying domain-specific syntax rules tailored to optical networking terminology and constraint patterns. Additionally, it enables error classification by distinguishing between issues that can be resolved automatically by the LLM and those that require user clarification. By acting as a structural verifier, the CFG ensures that only well-formed inputs reach downstream formal methods, thereby improving the reliability and interpretability of the entire pipeline.

\subsubsection{Optical RAG for Domain Knowledge Integration}

Generic LLMs lack the specialized domain knowledge necessary for accurate optical network design, such as ITU-T and IEEE standards, equipment specifications, and regulatory requirements. To address this, our Optical RAG component supplements the system with dynamic access to authoritative sources, enabling it to infer technical constraints, validate compliance with industry standards, and ensure equipment compatibility across vendors. 

Drawing inspiration from TelcoRAG~\cite{bornea2024telcoragnavigatingchallengesretrievalaugmented}, we adapted their domain-specific retrieval approach from 3GPP telecommunications standards to optical networking documentation, including ITU-T recommendations, IEEE optical standards, and vendor equipment specifications. By enriching user intents with structured, domain-specific guidance, Optical RAG is meant to enhance the technical soundness and realism of the generated designs.

\subsubsection{PDDL Planning for Formal Deployment Verification}

While CFG ensures structural validity and Optical RAG contributes domain expertise, neither alone can verify that a proposed network design is actually deployable. To address this, we incorporate formal semantic validation using the PDDL. This enables the system to verify deployment feasibility, enforce correct sequencing of actions, and validate resource constraints related to budget, timelines, and equipment availability. PDDL also ensures that the intended end-state of the network is reachable under these constraints.

\subsection{Three-Stage Pipeline Architecture}

\subsubsection{Stage 1: Intent Parser \& Validation}

\begin{figure}
    \centering
    \includegraphics[width=0.5\textwidth]{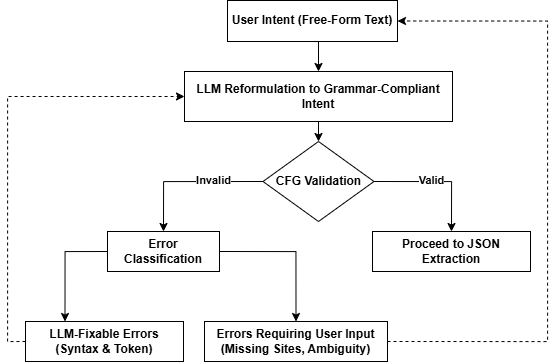}
    \caption{CFG parsing and validation process with intelligent error classification and routing}
    \label{fig:cfg-parsing}
\end{figure}

The first stage transforms natural language intents into validated structured representations. As detailed in Figure~\ref{fig:cfg-parsing}, the process begins with LLM-based translation to a grammar-compliant format, followed by CFG structural validation. 

We provide the LLM with the explicit grammar structure and instruct it to rephrase user intents into a compliant form. Our grammar enforces the following pattern:

\begin{tcolorbox}[colback=gray!10!white,boxrule=0.5pt,arc=2pt]
\footnotesize
\textbf{Core Grammar Structure:}
\begin{gather*}
\text{intent} \rightarrow \text{"We need a"} \; [\text{availability}] \; \text{"optical network connecting"} \; \\ \text{sites} \; [\text{constraints}] \\
\text{sites} \rightarrow \text{site} \; (\text{","} \; \text{site})^* \; \text{"and"} \; \text{site} \\
\text{site} \rightarrow \text{SITE\_NAME} \; [\text{location}] \; [\text{role}] \\
\text{role} \rightarrow \text{"("} \; (\text{"hub"} \mid \text{"core"} \mid \text{"edge"}) \; \text{")"} \\
\text{constraints} \rightarrow \text{constraint}^+ \\
\text{constraint} \rightarrow \text{latency} \mid \text{budget} \mid \text{compliance} \\
\text{budget} \rightarrow \text{"Our total budget for components is"} \; \text{VALID\_DOLLAR} \\
\text{VALID\_DOLLAR} \rightarrow \text{"\$"} \; \text{INT} \quad 
\end{gather*}
\end{tcolorbox}

This forces the LLM to fit user intents into predefined structural cases, creating a systematic constraint on LLM behavior. When information is missing from the user intent, the LLM faces a forced choice: either hallucinate the missing information or leave explicit gaps. If the LLM correctly avoids hallucination and omits unclear parameters, the grammar catches these omissions as structural violations. Conversely, if the LLM hallucinations occur—such as preserving vague terms like "fair" for budget constraints despite explicit instructions to use numeric values—the grammar's strict terminal validation rejects these as invalid format violations. This ensures that both missing information and LLM assumptions are systematically detected and flagged for user clarification, transforming the CFG from purely syntactic verification to also catch semantic gaps and common LLM errors.

Our error handling system classifies parsing failures into two categories. The first, \textit{LLM-fixable} errors, includes syntax and token formatting issues that can be resolved through automated self-correction. In these cases, the system automatically re-prompts the LLM with the original user intent and an explicit correction hint derived from the CFG error.

The second, \textit{user-required} errors, encompasses semantic ambiguities and missing information that require human clarification.
This classification enables appropriate routing of error correction, improving system efficiency compared to generic retry mechanisms. The stage outputs include structured intent JSON and formalized constraint specifications.

\subsubsection{Stage 2: Formal Planning \& Design}

\begin{figure*}[t]
    \centering
    \includegraphics[width=0.9\textwidth]{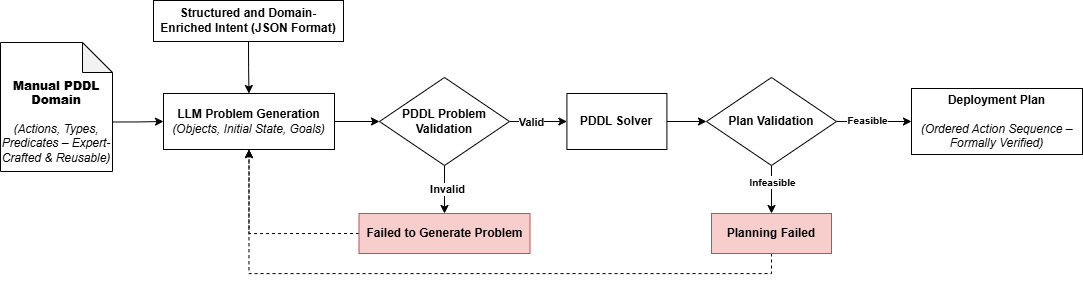}
    \caption{Hybrid PDDL planning approach combining manual domain expertise with LLM problem generation}
    \label{fig:pddl-planning}
\end{figure*}

The second stage enriches validated intents with domain knowledge and generates formal deployment plans. Our Optical RAG component integrates technical standards and equipment specifications into the structured intent from Stage 1, adapting RAG techniques for optical networking standards and vendor-specific documentation. This enriched intent then feeds into our hybrid PDDL approach (illustrated in Figure~\ref{fig:pddl-planning}) that combines expert-crafted domain knowledge with LLM-generated problem instances.

\textbf{Optical RAG Enhancement.} The Optical RAG system queries authoritative sources including ITU-T recommendations, IEEE optical networking standards, and vendor equipment specifications to inject domain-specific guidance into the structured intent. For example, when the user specifies "high-availability" connections, the RAG system retrieves relevant protection standards and adds specific guidance such as "implement 1+1 fiber protection" or "ensure geographically diverse routing." This enriched intent provides the technical context necessary for accurate PDDL problem generation and realistic network design.

\textbf{Hybrid PDDL Architecture.} As shown in Figure~\ref{fig:pddl-planning}, our approach separates static domain knowledge from dynamic, instance-specific input. The PDDL domain is manually crafted by experts and encodes the fundamental “physics” of optical network deployment—defining object types (e.g., sites, fibers, equipment), predicates (e.g., \texttt{site-operational}, \texttt{fiber-deployed}, \texttt{within-budget}), and actions (e.g., \texttt{commission-site}, \texttt{deploy-fiber}) along with their preconditions and effects. This domain remains fixed throughout the pipeline, capturing invariant rules and deployment logic.

Based on the enriched intent, the LLM generates a corresponding PDDL problem instance specifying the actual sites, equipment, initial state (e.g., infrastructure, budget), and goals (e.g., operational status, compliance). Constraints such as \texttt{"latency": 10ms} are translated into goal predicates like \texttt{(latency-satisfied)}, and RAG-derived recommendations are included in the formulation.

The PDDL solver then processes both components to generate a formal deployment plan—a sequence of grounded actions that transforms the initial state into the goal state while respecting all domain constraints. If the problem is unsolvable (e.g., conflicting constraints or physical impossibilities), the solver reports infeasibility, enabling explicit feedback to the user about why their requirements cannot be satisfied.

Our PDDL integration reduces error surface by having the LLM generate only problem instances rather than complete domains, enables reuse of expert-encoded planning knowledge across different scenarios, provides mathematical verification of plan correctness and optimality, and delivers explainable results each deployment action. The stage produces both deployment plans as ordered action sequences and network topology specifications that are formally verified for feasibility.

\subsubsection{Stage 3: Design Translation \& Validation}

The final stage employs an LLM agent to translate formal PDDL planning results into comprehensive, human-readable network designs and validates them against real-world constraints. 

\textbf{PDDL-to-Design Translation.} The LLM agent takes the formal deployment plan (sequence of grounded PDDL actions), enriched intent, and domain guidance as input to generate detailed network specifications. The agent interprets abstract PDDL actions like \texttt{(deploy-fiber SITE1 SITE2 fiber1)} into concrete engineering specifications including fiber types, equipment models, installation procedures, and cost estimates. This translation bridges the gap between formal planning abstractions and practical implementation details needed by network engineers.

\textbf{Comprehensive Design Output.} The pipeline produces complete network specifications including physical infrastructure (fiber routes, cable specifications), equipment requirements (ROADMs, amplifiers, transponders with specific models and costs), performance metrics (latency calculations, availability targets), cost breakdowns (infrastructure, equipment, installation labor), and implementation timelines aligned with the PDDL deployment sequence.

\textbf{Validation and Optimization.} Final validation ensures designs meet all specified constraints while providing optimization recommendations. When PDDL planning fails due to constraint conflicts, the system provides explicit feedback about infeasibility and suggests alternative approaches. The output maintains full traceability from user intent through formal planning to final design specifications.

\section{Validation and Case Studies}
\label{sec:eval}

We validate our pipeline through a proof-of-concept implementation and demonstrate its capabilities across representative scenarios. Our validation focuses on architectural completeness, component integration, and recovery mechanisms rather than large-scale evaluation.

\subsection{Implementation Overview}

Our prototype includes all three pipeline stages with error handling and recovery mechanisms. It is implemented in Python 3.9+ using OpenAI's GPT-4o-mini for LLM processing, Lark v1.1.5 for CFG parsing with LALR(1) grammar, and a custom PDDL integration combining manual domain specification with LLM-generated problem instances. The prototype features a graphical user interface for interactive testing and evaluation.

\subsection{Representative Case Studies}

\subsubsection{Case Study 1 - Complete Pipeline Success}

We use the following input intent: "\textit{We need a high-availability optical network connecting SITE1 (core), SITE2 (edge) and SITE3 (hub) support continuous operation with at least 3 geographically disjoint fiber paths between each pair of sites Maximum acceptable latency per path is 10 milliseconds Our total budget for components is \$1500000}"

\textbf{Pipeline Execution Flow:}

\textbf{Stage 1 - CFG-Guided Translation (245ms processing time):} The system provides the LLM with explicit grammar structure and instructs it to rephrase the complex natural language input. The LLM successfully maintains all site specifications, role assignments, and constraint requirements without hallucination or omission, producing a grammar-compliant structured sentence.

\textbf{Stage 2 - CFG Validation (12ms processing time):} Successfully parses and extracts components in a single pass:
\begin{itemize}
    \item Availability: "high-availability"
    \item Sites: SITE1 (core), SITE2 (edge), SITE3 (hub)
    \item Constraints: 3 disjoint paths, 10ms latency, \$1,500,000 budget
\end{itemize}

\textbf{Stage 3 - Optical RAG Enhancement (7.2s processing time):} Queries ITU-T recommendations for high-availability requirements and injects specific guidance: "Implement 1+1 fiber protection for critical sites," "Ensure geographic diversity for infrastructure resilience," "Use OS2 single-mode fiber for long-haul connections."

\textbf{Stage 4 - PDDL Planning (892ms processing time):} Manual domain defines optical deployment "physics" with 12 predicates and 8 actions including budget constraints. LLM generates problem instance with 3 sites, \$1,500,000 budget limit, and connectivity goals. PDDL solver validates resource constraints and produces 16-step deployment plan:
\begin{itemize}
    \item Steps 1-3: Commission all three sites
    \item Steps 4-6: Install ROADM equipment at each location (budget check: \$390K allocated)
    \item Steps 7-9: Deploy fiber optic cables between all site pairs (budget check: \$765K cumulative)
    \item Steps 10-12: Activate and configure ROADM equipment
    \item Steps 13-15: Activate fiber paths between all connections
    \item Step 16: Complete network deployment (final budget validation: \$1,400K within \$1,500K constraint)
\end{itemize}

\textbf{Stage 5 - Design Translation (2.1s):} LLM agent translates PDDL planning results into comprehensive network design, incorporating formal deployment sequence and domain guidance. Produces validated topology with \$1,400,000 cost estimate (within budget), 18-week implementation timeline, and complete equipment specifications.

The complete pipeline execution demonstrates strong performance across all validation metrics:
\begin{itemize}
    \item \textbf{Total Processing Time:} 10.3s
    \item \textbf{Pipeline Success Rate:} 100\% (all stages completed)
    \item \textbf{Constraint Satisfaction:} 100\% (budget, latency, redundancy met)
    \item \textbf{Formal Verification:} $\checkmark$ PDDL plan mathematically verified
\end{itemize}

The results illustrate the end-to-end hybrid pipeline effectiveness. They show that CFG prevents LLM assumptions while preserving intent and demonstrate PDDL formal verification with complex multi-constraint scenarios. Results also prove Optical RAG domain knowledge integration improves technical accuracy.

\subsubsection{Case Study 2 - Error Recovery and Quantitative Validation}
To demonstrate systematic error detection and recovery, we analyze this typical incomplete intent: "\textit{Build optical network with ROADM equipment and regulatory compliance}" (voluntarily missing site specification)

This intent demonstrates the system's error classification and recovery routing. The LLM successfully translates the input into a grammar-compliant format while preserving the missing site information. CFG validation detects \texttt{MissingSitesError} due to insufficient connectivity specification, which the system classifies as a "User-Required Error" requiring semantic clarification rather than automated correction. The system provides contextual guidance: "Please specify which sites/facilities you want to connect." After user correction, the pipeline completes successfully with a 3-site mesh topology.

To validate error recovery performance beyond individual examples, we conducted a comprehensive evaluation using a focused test corpus of 90 intent samples across three complexity levels (basic, intermediate, complex) for valid intents and four core error types for invalid intents. The corpus includes 30 correct intents designed to test grammar parsing capabilities and 60 incorrect intents (15 per error type) specifically crafted to trigger \texttt{MissingSitesError}, \texttt{VagueValueError}, \texttt{InvalidRoleError}, and \texttt{InvalidComplianceError} scenarios.

\begin{table}[ht]
\small
\centering
\caption{CFG Parsing Performance Evaluation}
\label{tab:cfg_performance}
\begin{tabular}{|l|c|}
\hline
\textbf{Metric} & \textbf{Result} \\
\hline
Valid Intent Pass-Through Rate & 96.7\% \\
Error Detection Rate & 96.7\% \\
Error Classification Accuracy & 100\% \\
Average Processing Time & 710.1ms \\
\hline
\end{tabular}
\end{table}

\begin{table}[ht]
\small
\centering
\caption{Error Type-Specific Performance}
\label{tab:error_performance}
\begin{tabular}{|l|c|c|c|c|}
\hline
\textbf{Error Type} & \textbf{Tests} & \textbf{Detected} & \textbf{Correct} & \textbf{Acc.\%} \\
\hline
MissingSites & 15 & 15 & 15 & 100\% \\
VagueValue & 15 & 13 & 13 & 86.7\% \\
InvalidRole & 15 & 15 & 15 & 100\% \\
InvalidCompliance & 15 & 15 & 15 & 100\% \\
\hline
\end{tabular}
\end{table}

The evaluation results are presented in Tables~\ref{tab:cfg_performance} and~\ref{tab:error_performance}, measuring system performance using the following metrics:

\begin{itemize}
    \item \textbf{Valid Intent Pass-Through Rate}: Percentage of correctly formed intents (30 test cases) that successfully complete CFG parsing without triggering false error detection.
    \item \textbf{Error Detection Rate}: Percentage of intentionally malformed intents (60 test cases) that are correctly identified as containing structural or semantic errors.
    \item \textbf{Error Classification Accuracy}: Among successfully detected errors, percentage that are correctly categorized into the appropriate error type.
\end{itemize}

The results demonstrate good performance across most error categories. Perfect detection and classification rates for \texttt{MissingSitesError}, \texttt{InvalidRoleError}, and \texttt{InvalidComplianceError} validate the effectiveness of our CFG terminal symbol validation and semantic analysis approaches. The 86.7\% detection rate for \texttt{VagueValueError} reflects a specific LLM behavior: when encountering vague terms like "reasonable latency" or "several paths," the LLM sometimes chooses to omit the entire constraint rather than preserve the vague value in its grammar-compliant output. This constraint omission results in syntactically valid sentences that do not trigger CFG validation errors, causing 2 out of 15 VagueValueError cases to pass undetected. However, all detected vague value cases (13/13) were correctly classified, demonstrating 100\% classification accuracy. 

For valid intents, the CFG parsing achieved a 96.7\% pass-through rate overall: 10/10 for basic cases, 10/10 for intermediate cases, and 9/10 for complex cases. This confirms that the system processes well-formed specifications reliably without false rejections, even at higher complexity levels.

The evaluation results confirm that the recovery behaviors observed in case studies are systematic and repeatable. Notably, the \texttt{VagueValueError} results highlight an architectural strength: grammar-guided translation forces the LLM into structured choices that expose uncertainty rather than allowing hallucinated precision. When encountering ambiguous terms, the constrained LLM either produces valid numeric values or omits constraints entirely, both of which our system can handle appropriately.

\subsubsection{Case Study 3 - Formal Constraint Analysis and Graceful Degradation}
This case study demonstrates the system's ability to detect and handle physically impossible requirements through formal verification. When presented with the intent "\textit{Connect 15 sites across continental US with sub-millisecond latency}" the system successfully parses the request through CFG validation but identifies fundamental physics violations during PDDL planning. The formal solver determines that sub-millisecond latency is impossible over continental distances given the speed of light in fiber (~200,000 km/s), providing mathematical proof of infeasibility rather than generating plausible but incorrect designs. Upon detecting unsatisfiable constraints, the system gracefully degrades to LLM-only topology generation while explicitly disclosing the limitation and providing educational feedback about physical constraints. This approach contrasts sharply with pure LLM systems that might generate technically impossible but superficially reasonable network designs, demonstrating the value of formal verification in preventing silent failures and educating users about real-world limitations.

\subsection{Discussion}
The three case studies demonstrate the system’s capability to address representative intent processing scenarios encountered in practice. Case 1 validates the complete pipeline functionality with complex, well-specified intents representing the 96.7\% of valid inputs successfully processed in our evaluation corpus. Case 2 showcases error classification and recovery mechanisms for the 96.7\% of problematic intents that the system correctly detects, exemplified through \texttt{MissingSitesError} handling that guides users toward specification completion. Case 3 highlights the critical value of formal verification in identifying physically impossible requirements that pure LLM approaches would miss, providing educational feedback about real-world constraints rather than generating plausible but incorrect designs. 
Together, these cases validate our quantitative evaluation
results while demonstrating the practical benefits of the hybrid
architecture across diverse real-world scenarios.
\section{Conclusion and Future Perspectives}
\label{sec:concl}

We presented a hybrid pipeline that integrates formal methods with LLMs to translate natural language intents into verifiable optical network designs. By combining CFG parsing, domain-informed RAG, and PDDL planning, the system balances accessibility with formal correctness.

The CFG parsing component, evaluated on 90 diverse intents, achieves a 96.7\% pass-through rate for valid inputs, 96.7\% error detection, and 100\% classification accuracy. Complete pipeline validation through representative case studies demonstrates the pipeline’s ability to handle complex multi-constraint designs and degrade gracefully when requirements are infeasible, offering educational feedback rather than silent failure.

Future work includes optimizing runtime (particularly in RAG and PDDL stages), expanding domain support to more complex optical technologies, and integrating multi-objective planning with adaptive learning while maintaining formal guarantees. Beyond design, extending the pipeline toward full IBN workflows—including configuration synthesis, intent refinement, and closed-loop automation—could enable end-to-end network deployment and full lifecycle management.


\bibliographystyle{IEEEtran}
\bibliography{references}

\end{document}